\documentclass[amsmath,amssymb, 12pt]{revtex4}
\usepackage{graphicx}
\usepackage{dcolumn}
\usepackage{bm}
\usepackage{graphicx}
\usepackage{epsfig}

\def\bea{\begin{eqnarray}}
\def\eea{\end{eqnarray}}

\begin{document}
\title{Dark energy interacting with two fluids}
\author{Norman Cruz}
\altaffiliation{ncruz@lauca.usach.cl} \affiliation{Departamento de
F\'\i sica, Facultad de Ciencia, Universidad de Santiago, Casilla
307, Santiago, Chile.}
\author{Samuel Lepe}
\altaffiliation{slepe@ucv.cl} \affiliation{Instituto de F\'\i
sica, Facultad de Ciencias B\'asicas y Matem\'aticas, Universidad
Cat\'olica de Valpara\'\i so, Avenida Brasil 2950, Valpara\'\i so,
Chile.}
\author{Francisco Pe\~na}
\altaffiliation{fcampos@ufro.cl} \affiliation{Departamento de
Ciencias F\'\i sicas, Facultad de Ingenier\'\i a, Ciencias y
Administraci\'on, Universidad de La Frontera, Avda. Francisco
Salazar 01145, Casilla 54-D Temuco, Chile.\\}
\date{\today}
\begin{abstract}
A cosmological model of dark energy interacting with dark matter
and another general component of the universe is investigated. We
found general constraints on these models imposing an accelerated
expansion. The same is also studied in the case for holographic
dark energy.
\end{abstract}
\maketitle
\section{ Introduction}

The existence of a dark component with an exotic equation of state
~\cite{Spergel} filling a flat universe~\cite{Jae} has been well
established from the combined MAXIMA-1~\cite{Hanany},
BOOMERANG~\cite{Mauskopf}, DASI~\cite{Halverson} and COBE/DMR
Cosmic Microwave Background (CMB)~\cite{Smoot} observations. These
results have been confirmed and improved by the recent WMAP
data~\cite{Spergel1}. The accelerated expansion is consistent
with the luminosity distance as a function of redshift of distant
supernovae~\cite{Riess}, the structure formation
(LSS)~\cite{Percival} and the cosmic microwave background
(CMB)~\cite{D.N.Spergel}. This dark energy constitute the
component with the biggest contribution to the energy density,
responsible for $70$ of the total energy density and has an
equation of state similar to that of a cosmological constant, i.
e., with a ratio $w =p/\rho$ negative and close to -1.
Additionally, the other important contribution to the total
density is the dark matter, which is roughly 1/3 of the total
energy density, and is a gravitationally interacting form of non
baryonic matter. Its existence have been well established by the
observations of rotation curves in galaxies~\cite{Feldman} and
the CMB observations by WMAP~\cite{Spergel},~\cite{D.N.Spergel}.

The cosmic observations show that densities of dark energy and
dark matter are of the same order today. To solve this coincidence
problem~\cite{Weinberg} (or why we are accelerating in the current
epoch due that the vacuum and dust energy density are of the same
order today ?) it is assumed an evolving dark energy field with a
non-gravitational interaction with matter~\cite{Amendola} (decay
of dark energy to matter). In the case of the coupling between
matter and quintessence it is motivated for string theory or
arises after a conformal transformation of Brans-Dicke
theory~\cite{Amendola1}. In general, it is assumed that the dark
matter and dark energy are coupled by a term Q that gauges the
transfer of energy from the dark energy to the dark matter.
Interacting quintessence model, with a Q term proportional to the
sum dark matter density and dark energy density, has been tested
studying the effect of the coupling in the matter power spectrum,
constraining the parameters of the models using the matter power
spectrum measured by the 2-degree fiel galaxy redshift survey
2dFGRS~\cite{Olivares}. The consequences of an interaction
between dark matter and dark energy on the rates for the direct
detection of dark matter were investigated at galactic level
in~\cite{Tetradis}. In this case the interaction was modeled by
assuming that the mass of the dark matter depends on the scalar
field whose potential provides the dark energy. Very recently,
the Abel cluster was investigated as an example of self
gravitating  system where interaction  between dark energy and
dark matter can be detected~\cite{Bertolami}.

In the above discussion we have centered in the interaction
between dark energy and dark matter. Nevertheless, it is
physically reasonable and even expected from a theoretical point
of view, that dark matter as well as dark energy can interact with
other components of the universe. See, for
example,~\cite{Brokfield} for a model of dark energy interacting
with neutrinos. If we focus in the possible decaying of dark
energy, $w < -1$ is an indication that dark energy does indeed
interact with another fluid. In~\cite{Macorra} was studied the
conditions under a dark energy can dilute faster or decay into
the fermion fields. Kremer~\cite{Kremer} has investigated a
phenomenological theory of dark energy, with a given equation of
state, interacting with neutrinos and dark matter.

Since a complete understanding of the nature of dark energy may
involve to formulate a consistent theory of quantum gravity, not
yet found, we make some attempts to clarify the nature of this
energy following one of the new principles that emerges as a
guideline to this ultimate theory, which is the holographic
principle. This principle says that the number of degrees of
freedom of a physical system should scale with its bounding area
rather than with its volume. In the holographic dark energy
models the investigations have been made in order to explain the
size of the dark energy density on the basis of holographic
ideas, derived from the suggestions that in quantum field theory a
short distance cut-off is related to a long distance cut-off due
to the limit set by the formation of a black hole~\cite{Cohen}.
In~\cite{Li} the future event horizon was considered as the long
distance in order to recover the equation of state for a dark
energy dominated universe.

Following the previous results exposed above about the
interactions between dark energy and dark matter and another
component of the universe, our aim in the present work is to
investigate the general constraints on a cosmological model where
the interaction between dark energy, dark matter and another
fluid, which we can not identified specifically, is considered.
The will deduce restrictions for these models when the
accelerated expansion is imposed and when, additionally, an
holographic dark energy is taken account, which assume the
following form
\begin{eqnarray}
\label{eq1} \rho_{\chi}=3c^{2}M_{P}^{2}L^{-2}
\end{eqnarray}
where $c$ is a numerical constant, $M_{P}=\frac{1}{\sqrt{8\pi
G}}$ is the reduced Planck mass and $L$ is a characteristic
infrared cutoff. In this work we shall identified the Hubble
horizon as $L$.

Our paper is organized as follows.  In section I we present the
model for a universe filled with dark matter, dark energy and
another fluid. We shall impose that the interacting term $Q$ and
$Q'$ which appears in the conservation equations are different.
We derive general constraints for the difference between $Q$ and
$Q'$ imposing an accelerated expansion. In section II we discuss
this cosmological scenario with an holographic dark energy and
other constraints are presented. In section III we briefly discuss
our results.

\section{Interacting dark energy}

In the following we modeled the universe made of CDM with a
density $\rho_{m}$ and a dark energy component, $\rho_{DE}$, which
obey the holographic principle. We will assume that the the dark
matter component is interacting with the dark energy component,
so their continuity equations take the form
\begin{eqnarray}\label{continuity1}
\overset{\cdot }{\rho }_{DE}+3H\left( \rho _{DE}+p_{DE}\right)
&=&-Q^{\prime},
\end{eqnarray}
\begin{eqnarray}\label{continuity2}
\overset{\cdot }{\rho }_{m}+3H\left( \rho _{m}+p_{m}\right) &=&Q.
\end{eqnarray}
Note that we have wrote $Q$ and $Q^{\prime}$ in order to include
the scenario in which the mutual interaction between the two
principal components of the universe leads to some loss in other
forms of cosmic constituents. In this case $Q \neq Q^{\prime}$.
Note we do not set a priori the sign of $Q$ and $Q \neq
Q^{\prime}$, which means that we obtain the corresponding sign of
this quantities from the constraints that we obtain bellow. If we
denote this other component by $\rho_{X}$, its corresponding
continuity equation is given by
\begin{eqnarray}\label{continuity3}
\overset{\cdot }{\rho }_{X}+3H\left( \rho _{X}+p_{X}\right)
&=&Q^{\prime}-Q.
\end{eqnarray}
We are taking about in this case that dark energy decay into dark
matter (or viceversa, depending on the sign of Q) and other
component. Of course, since our aim is to shed some light on the
coincidence problem and the holographic principle, we expect to
have $Q^{\prime}-Q \ll 1$. The sourced Friedmann equation is then
given by
\begin{eqnarray}\label{sourcedFridman}
3H^{2} &=&\rho _{DE}+\rho _{m}+\rho_{X}-\frac{3k}{a^2}.
\end{eqnarray}
In the following we assume that $\rho _{DE}\gg \rho_{X}$ and
$\rho _{m}\gg \rho_{X}$, which is consistent with the observable
content of the universe, so $\rho_{X}$ can be neglected from
Eq.~(\ref{sourcedFridman}). Assuming that the equations of state
for dark matter and dark energy are given by $p_{i}=\omega
_{i}\rho _{i}$, where $\omega _{i}=\omega _{i}\left(H\right)$
$(i=1,2)$, Eq.~(\ref{continuity1}) and Eq.~(\ref{continuity2})
can be written only in terms of the density
\begin{eqnarray}\label{continuity11}
\overset{\cdot }{\rho }_{DE}+3\left(1+
\omega_{DE}\right)H\rho_{DE} &=&-Q^{\prime},
\end{eqnarray}
\begin{eqnarray}\label{continuity22}
\overset{\cdot }{\rho }_{m}+3H\left[ (1+\omega_{m})\rho_{m}+\pi
\right] &=&Q^{\prime}.
\end{eqnarray}
Adding Eq.~(\ref{continuity11}) and Eq.~(\ref{continuity22}) we
obtain
\begin{eqnarray}\label{continuityrho}
\dot{\rho}+3[(1+\omega)\rho+\pi]H=0,
\end{eqnarray}
where $\rho$, $\omega$ and $\pi$ are defined by
\begin{eqnarray}\label{rhoequiv}
\rho\equiv\rho_{DE}+\rho_{m},
\end{eqnarray}
\begin{eqnarray}\label{omegaequiv}
\omega \rho\equiv \omega_{DE}\rho_{DE}+\omega_{m}\rho_{m},
\end{eqnarray}
and
\begin{eqnarray}\label{pi}
\pi=-\frac{Q-Q^{\prime}}{3H}.
\end{eqnarray}
Using Eqs.~(\ref{continuityrho}) and~(\ref{sourcedFridman}) and
the definitions given by
Eqs.~(\ref{rhoequiv}),~(\ref{omegaequiv}) and~(\ref{pi}), we
obtain
\begin{equation}\label{eqforomega}
1+\omega=-\frac{1}{H^2+\frac{k}{a^2}}\left(\frac{2}{3}({\dot{H}-
\frac{k}{a^{2}}})+\frac{\pi}{3}\right).
\end{equation}
Let us first consider the constraint for $\pi$ and $Q$ derived
from the condition of accelerated universe. The second Friedmann
equation is
\begin{eqnarray}\label{secondFriedman}
2\dot{H}+3H^{2}+\frac{k}{a^{2}}=-(p_{DE}+p_{m}+\pi).
\end{eqnarray}
Using Eq.~(\ref{sourcedFridman}) in Eq.~(\ref{secondFriedman})
yields
\begin{eqnarray}
\dot{H}=-\frac{1}{2}(1+\omega_{DE}+(1+\omega_{m})r+
\frac{\pi}{\rho_{DE}})\rho_{DE}+\frac{k}{a^{2}}.
\end{eqnarray}
The expression for the acceleration becomes
\begin{eqnarray}
\frac{\ddot{a}}{a}=-\frac{1}{6}(1+3\,\omega_{DE}+(1+3\,\omega_{m})r+3\frac{\pi}{\rho_{DE}})\rho_{DE},
\end{eqnarray}
where $r=\frac{\rho_{m}}{\rho_{DE}}$ is the ratio between dark
matter and dark energy. For an accelerated universe,
$\ddot{a}>0$, the parameter $\pi$ must satisfy the following
inequality, which is independent of $k$,
\begin{eqnarray}\label{ineqforpi}
\pi<-\frac{1}{3}[(1+3\,\omega_{DE})+(1+3\,\omega_{m})r]\rho_{DE}.
\end{eqnarray}
If we evaluate $\pi$ at the present time, taking $r\approx 0.42$
and $\omega_{m}=0$, we obtain
\begin{eqnarray}\label{ineqforpi0}
\pi(t=0)<-\frac{1}{3}[1.42+3\,\omega_{DE}]\rho_{DE}.
\end{eqnarray}
We have two possible scenarios: In the first one
$1.42+3\,\omega_{DE}<0$, which implies that $\pi$ must be lower
than a positive value. Then it is possible to take $\pi >0$, or
in other words $Q<Q^{\prime}$, i.e., part of the dark energy
density decays in dark matter and the rest in the other unknown
energy density component, $\rho_{x}$. In this case
$\omega_{DE}<-0.47$. In the second one $1.42+3\,\omega_{DE}>0$,
which implies that $\pi<0$ or equivalently $Q>Q^{\prime}$. In
this case the dark matter received energy from the dark energy
and from the unknown component and $\omega_{DE}>-0.47$. Obviously,
observations favored the first scenario with decaying dark energy
into dark matter.

If we assume that the unknown mechanism which leads to an
interaction between dark energy and dark matter always implies
$\pi>0$ during the cosmic evolution, then in the non accelerated
phase the Eq.~(\ref{ineqforpi}) with $\omega_{m}=0$ becomes
\begin{eqnarray}\label{ineqforpi1}
\pi(t) >-\frac{1}{3}[(1+3\,\omega_{DE}(t))+r(t)]\rho_{DE}.
\end{eqnarray}
Since in our model the density of dark energy is a function of
time we write $\omega_{DE}(t)$. Note that in our general approach
we not assume a specific model for the dark energy and the
interaction $Q$. In this case, the expression
$1+3\,\omega_{DE}(t)+r(t)$ could be positive or negative. If it
positive no new constraint is obtained for $\pi$. If it is
negative we obtain a lower value for $\pi$.

Since we have not imposed a specific models for the dark energy
and the interaction terms $Q$ and $Q^{\prime}$, we do no have any
parameters that can constrained by observations.

\section{Holographic dark energy}

In the following we consider that the dark energy is given by
\begin{eqnarray}\label{holoenergy}
\rho_{DE}=3c^{2}H^{2}.
\end{eqnarray}
It is known that the without interaction between the two
principal components in a flat universe $\rho_{DE}$ behaves as
$\rho_{m}$ because of $\rho_{m} \sim H^{2} \sim \rho_{DE}$
~\cite{Li}. Let us discuss a general result which involves the
curvature of the universe, in the framework of a holographic dark
energy, filled with three interacting fluids, which implies $\pi
\neq 0$, i.e., $Q\neq Q^{\prime}$. The Eq.~(\ref{sourcedFridman})
can be rewritten as
\begin{eqnarray}\label{sourcedFridman1}
1=\frac{\rho_{DE}}{3H^{2}}(1+r+s)+\Omega_{k},
\end{eqnarray}
where $s=\rho_{X}/\rho_{DE}$ and $\Omega_{k}\equiv
-k/a^{2}H^{2}$. Using Eq.~(\ref{holoenergy}) in
Eq.~(\ref{sourcedFridman1}) we obtain
\begin{eqnarray}\label{sourcedFridman2}
1=c^{2}(1+r+s)+\Omega_{k}.
\end{eqnarray}
For a flat universe the above equation tell us that $r+s$ must be
constant, i.e., $\dot{r}+\dot{s}=0$. If we assume the standard
approach $Q=Q^{\prime}$, an holographic dark energy implies that
$r=constant$ for a flat universe. It is clear that depending, for
example, in the choice that we can made to model $Q$, $r$ can be
variable during the cosmic evolution, obtaining a more suitable
model to describe the cosmic coincidence.

In the following we discuss the contribution of $\pi$ in order to
obtain more negative $\omega_{DE}$ for the equation of state of
the dark energy. For a flat universe with interaction
Eq.~(\ref{sourcedFridman}) becomes (taking into account that
$\rho _{DE}\gg \rho_{X}$ and $\rho _{m}\gg \rho_{X}$)
\begin{eqnarray}\label{matterholo}
\rho_{m}=3(1-c^{2})H^{2}.
\end{eqnarray}
Deriving the above equation and using Eq.~(\ref{continuity2})
yields
\begin{eqnarray}\label{hh1}
\frac{2}{3}\frac{\dot{H}}{H^{2}}=\frac{Q}{9(1-c^{2})H^{3}}-(1+\omega_{m}).
\end{eqnarray}
For a flat universe Eq.~(\ref{eqforomega}) becomes
\begin{eqnarray}\label{hh2}
\frac{2}{3}\frac{\dot{H}}{H^{2}}=\frac{\pi}{3H^{2}}-(1+\omega).
\end{eqnarray}
Comparing Eq.~(\ref{hh1}) and Eq.~(\ref{hh2}) we obtain
\begin{eqnarray}\label{hh3}
\omega =
\omega_{m}-\left[\frac{Q}{9(1-c^{2})H^{3}}+\frac{\pi}{3H^{2}}
\right],
\end{eqnarray}
which means that if the matter content is described by dust,
i.e., $\omega_{m}=0$, and Hubble horizon as a characteristic
length, the equivalent equation of the state for a universe with
dark energy is different from zero. This result is obtained when
exist $Q=Q^{\prime}\neq 0$.  In our case, a positive $\pi$, i.e.,
$Q<Q^{\prime}\neq 0$, helps to a more negative value for $\omega$.
For a non flat universe the equation of state is given by
\begin{eqnarray}\label{hh7}
1+\omega=\frac{1}{1+k(aH)^{-2}}\left(1+ \omega_{m}- \frac{1}{3}
\left[\frac{Q}{3(1-c^{2})H^{2}}+\frac{\pi}{H^{2}}-2k(aH)^{-2}\right]\right)
\end{eqnarray}

Using the holographic dark energy we can obtain a new constraint
for $\pi$, which involves the parameter $c$. For an accelerated
universe, $\ddot{a}$, the constraint for $\pi$ in terms of
$\omega$ (defined by Eq.~(\ref{omegaequiv})), and $\rho$ (defined
by Eq.~(\ref{rhoequiv})) is given by
\begin{eqnarray}\label{piconstraint}
\pi < - \left(\omega + \frac{1}{3}\right)\rho,
\end{eqnarray}
which means that $\omega <- 1/3$ for a positive $\pi$. Using the
holographic condition given in Eq.~(\ref{holoenergy}) in
Eq.~(\ref{piconstraint}) in $\omega <- 1/3$, we obtain
\begin{eqnarray}\label{pionH}
\frac{\pi}{H^{2}} < 3c^{2}(|\omega| - \frac{1}{3})(1+r).
\end{eqnarray}
Evaluating the above inequality at the present time, where
$|\omega|\approx 1$ and $r_{0}\approx 3/7$ yields
\begin{eqnarray}\label{pionHincero}
\frac{\pi}{H^{2}}|_{0} < \frac{20}{7}c^{2}.
\end{eqnarray}
Since $c^{2}<1$, in order to have a non zero dark matter density
(see Eq.(\ref{matterholo}), the above equation indicates that
$\pi \sim H_{0}^{2}$.

\section{Discussion}

In the present investigation we have considered a cosmological
scenario where exist interactions between dark energy, dark
matter and another component of the universe, which we do not
identify explicitly. We have obtained general constraints,
considering an accelerated expansion, on the parameter $\pi$. In
the context of an holographic dark energy, and using the Hubble
radius as infrared cutoff, we have shown that our scenario leads
naturally, for a flat universe, to a more suitable approach to the
cosmic coincidence problem in which $r$ can be variable during
the cosmic evolution. We also found, in this context, that a
positive $\pi$, i.e., $Q<Q^{\prime}\neq 0$, helps to obtain a
more negative value for $\omega$, as can be seen from
Eq.(\ref{hh3}). Finally, a constraint for $\pi$ at the present
cosmic time as been found, which can be of the order of
$H_{0}^{2}$. This constraint is obtained in the framework our the
holographic approach and can be considered as only a first
approximation to obtain a range for this parameter.


\section{acknowledgements}
NC and SL acknowledge the hospitality of the Physics Department of
Universidad de La Frontera where part of this work was done. SL
acknowledges the hospitality of the Physics Department of
Universidad de Santiago de Chile.  We acknowledge the partial
support to this research by CONICYT through grant N$^0$ 1040229
(NC and SL). It also was supported from DIUFRO N$^0$ 120618, of
Direcci\'on de Investigaci\'on y Desarrollo, Universidad de La
Frontera (FP) and DI-PUCV, Grants 123.792/07, Pontificia
Universidad Cat\'olica de Valpara\'\i so (SL).


\begin{thebibliography}{xxxxx}

\bibitem{Spergel}D. N. Spergel et al. [WMAP Collaboration],
Astrophys. J. Suppl. 148 (2003) 175, [arXiv:astro-/0302209]. A.
\bibitem{Jae}H. Jae et al. [Boomerang Collaboration], Phys. Rev. Lett. 86
(2001) 3475, [arXiv:astro-ph/0007333].
\bibitem{Hanany}S. Hanany et al., Astrophys. J. 545 (2000) L5,
[arXiv:astro-ph/0005123]; R. Stompor et al.,Astrophys. J. 561
(2001) L7, [arXiv:astro-ph/0105062]; M. G. Santos et al., Phys.
Rev. Lett. 88 (2002) 241302, [arXiv:astro-ph/0107588].
\bibitem{Mauskopf}P. D. Mauskopf et al. [Boomerang Collaboration], Astrophys.
J. 536 (2000) L59, [arXiv:astro-h/9911444]; S. Masi et al., Prog.
Part. Nucl. Phys. 48 (2002) 243 [arXiv:astro-ph/0201137]; J. E.
Ruhl et al., Astrophys. J. 599 (2003) 786
[arXiv:astro-ph/0212229].
\bibitem{Halverson}N. W. Halverson et al., Astrophys. J. 568 (2002) 38
[arXiv:astro-ph/0104489]; J. L. Sievers et al., Astrophys. J. 591
(2003) 599 [arXiv:astro-ph/0205387].
\bibitem{Smoot}G. F. Smoot et al., Astrophys. J. 396 (1992) L1.
\bibitem{Spergel1}D. N. Spergel et al., [arXiv:astro-ph/0603449]; L. Page et
al., [arXiv:astro-ph/0603450]; G. Hinshaw et al.,
[arXiv:astro-ph/0603451]; N. Jarosik et al.,
[arXiv:astro-ph/0603452].
\bibitem{Riess}A. G. Riess et al. [Supernova Search Team Collaboration],
Astron. J. 116 (1998) 1009      [arXiv:astro-ph/9805201] ; S.
Perlmutter et al. [Supernova Cosmology Project Collaboration],
Astrophys. J. 517 (1999) 565, [arXiv:astro-ph/9812133]; W. J.
Percival et al. [The 2dFGRS Collaboration], Mon. Not. Roy.
Astron. Soc. 327 (2001) 1297, [arXiv:astro-ph/0105252]; P. Astier
et al., [arXiv:astro-ph/05 10 447] ; A. G. Riess et al.
[Supernova Search TeamCollaboration], Astrophys. J. 607, 665
(2004) [arXiv:astro-ph/0402512].
\bibitem{Percival}W. J. Percival et al. [The 2dFGRS Collaboration], Mon. Not.
Roy. Astron. Soc. 327, 1297 (2001) [arXiv:astro-ph/0105252]; M.
Tegmark et al. [SDSS Collaboration], Phys. Rev. D 69 (2004)103501 [arXiv:astro-ph/0310723].
\bibitem{D.N.Spergel}D. N. Spergel et al., [arXiv:astro-ph/0603449]; A. C. S.
Readhead et al., Astrophys. J. 609 (2004)498
[arXiv:astro-ph/0402359]; J. H. Goldstein et al., Astrophys. J.
599, 773 (2003) [arXiv:astro-ph/0212517].
\bibitem{Feldman}H. Feldman Astrophysical Journal 596 (2003) L131.
\bibitem{Weinberg}S. Weinberg, Rev. Mod. Phys. 61, 1 (1989); I. Zlatev, L.-M. Wang,
and P. J. Steinhardt, Phys. Rev.   Lett. 82, 896 (1999).
\bibitem{Amendola} L. Amendola, Phys. Rev. D 62, 043511 (2000);
W. Zimdahl, D. Pav\'on and L.P. Chimento, Phys. Lett. B 521, 133
(2001);M. Gasperini, F. Piazza and G. Veneziano, Phys. Rev. D65,
023508 (2002); W. Zimdahl, Int. J. Mod. Phys. D 14, 2319 (2005);
D. Pav\'on and W.Zimdahl, Phys.   Lett. B 628, 206 (2005); G.
Mangano, G. Miele and V. Pettorino, Mod. Phys. Lett.A 18, 831
2003); G. Farrar and P.J.E. Peebles, Astrophys. J. 604, 1 (2004);
S. del Campo, R. Herrera, and D. Pav\'on, Phys. Rev. D 70, 043540
(2004); R. Cai and A. Wang, J. Cosmol. Astropart. Phys.03, 002
(2005); Micheal S. Berger and H. Shojaei, Phys. Rev. D 73, 083528
(2006); Bo Hu and Y. Ling, Phys. Rev. D 73, 123510 (2006); Hui
Li, Z. Guo and Y. Zhang, Int. J. Mod. Phys. D 15, 869 (2006); A.
P. Billyard and A. A. Coley, Phys. Rev. D 61, 083503 (2000); M.
Szydlowski, Phys. Lett. B 632, 1 (2006); M. Szydlowski, T.
Stachowiak and R. Wojtak, Phys. Rev. D 73, 063516 (2006); L. P.
Chimento, A. S. Jakubi, D. Pav\'on and W. Zimdahl, Phys. Rev.D 67,
083513 (2003); L. P. Chimento and D. Pav\'on, Phys. Rev.D 73,
063511 (2006); G. Olivares, F. Atrio-Barandela, and D. Pav\'on,
[arXiv:gr-qc/0601086]; L. Amendola, G. Camargo Campos and R.
Rosenfeld [arXiv:gr-qc/061006].
\bibitem{Amendola1} L. Amendola, Phys. Rev. D 62, 043511 (2000);
L. Amendola and L. Tocchini-Valentini, Phys. Rev.   D 64, 043509
(2001); L. Amendola and D. Tocchini-Valentini, Phys. Rev. D 66,
043528 (2002); L. Amendola, C. Quercellini, D.
Tocchini-Valentini, and A. Pasqui, Astrophys. J. 583, L53(2003).
\bibitem{Olivares}G. Olivares, F. Atrio-Barandela, and D. Pav\'on,
[arXiv:gr-qc/0601086]; [arXiv:astro-ph/0612773].
\bibitem{Tetradis} N. Tetradis, J. D. Vergados and Amand Faessler,
Phys. Rev. D 75, 023504 (2007) [arXiv:astro-ph/0609078].
\bibitem{Bertolami} O. Bertolami , F. Gil Pedro and M. Le Delliou,
[arXiv:astro-ph/0703462].
\bibitem{Brokfield} A. W. Brokfield, C. van de Bruck, D. F. Mota
and D. Tocchini-Valentini, Phys. Rev. Lett. 96, 061301 (2006);
Phys. Rev. D 73 085515 (2006).
\bibitem{Macorra} A. de la Macorra,[arXiv:astro-ph/0702239];
[arXiv:astro-ph/0701635].
\bibitem{Kremer} G. M. Kremer,[arXiv:gr-qc/0704.0371].
\bibitem{Cohen} A. Cohen, D. Kaplan, and A. Nelson,
Phys. Rev. Lett. 82, 4971 (1999) [arXiv:hep-th/9803132].
\bibitem{Li} M. Li, Phys. Lett. B 603, 1 (2004)[hep-th/0403127].



\end{thebibliography}
\end{document}